\documentclass[11pt]{article}
\usepackage{moriond,epsfig}
\usepackage{multirow}

\def\be{\begin{equation}}
\def\ee{\end{equation}}
\def\bea{\begin{eqnarray}}
\def\eea{\end{eqnarray}}

\providecommand{\MT}{\ensuremath{M_{\mathrm{T}}\xspace}}

\newcommand{\mH}{\ensuremath{m_{\mathrm{H}}}\xspace}
\newcommand{\MW}{\ensuremath{m_{\mathrm{W}}}\xspace}

\RequirePackage{heppennames2}
\RequirePackage{xspace}
\RequirePackage{amsmath}

\newcommand{\MH}{\ensuremath{M_{\PH}}}
\newcommand{\fbinv} {\mbox{\ensuremath{\,\text{fb}^\text{$-$1}}}\xspace}
\newcommand{\GeV}{\ensuremath{\,\text{Ge\hspace{-.08em}V}}\xspace}

\newcommand{\Pgg}{\PGg}
\newcommand{\Pgt}{\PGt}
\newcommand{\cPZ}{\PZ} % plain Z (no superscript 0)
\newcommand{\cPqb}{\PQb} % b for b quark

\newcommand{\et}{\ensuremath{E_{\mathrm{T}}}\xspace}
\newcommand{\PT}{\ensuremath{p_{\mathrm{T}}}\xspace}
\newcommand{\pt}{\ensuremath{p_{\mathrm{T}}}\xspace}

\newcommand{\ETm}{\ensuremath{E_{\mathrm{T}}^{\text{miss}}}\xspace}
\newcommand{\MET}{\ETm}

\begin{document}
\vspace*{4cm}
\title{Searches for the Standard Model Scalar Boson at CMS}

\author{Marco Pieri for the CMS Collaboration}

\address{University of California San Diego, La Jolla, CA, U.S.A}

\maketitle\abstracts{
We searched for the standard model Higgs boson in many different channels using approximately 5 \fbinv of 7 TeV pp collisions data collected with the CMS detector at LHC. Combining the results of the different searches we exclude at 95\% confidence level a standard model Higgs boson with mass between 127.5 and 600 GeV. The expected  95\% confidence level exclusion if the Higgs boson is not present is from 114.5 and 543 GeV. The observed exclusion is weaker than expected at low mass because of some excess that is observed below about 128 GeV. The most significant excess is found at 125 GeV with a local significance of $2.8 \sigma$. It has a global significance of $0.8 \sigma$ when evaluated in the full search range and of $2.1 \sigma$ when evaluated in the range 110--145 GeV. The excess is consistent both with background fluctuation and a standard model Higgs boson with mass of about 125 GeV, and more data are needed to investigate its origin.
}

%%%%%%%%%%%%%%%%%
%
% Introduction
%
%%%%%%%%%%%%%%%%%

\section{Introduction}

The scalar boson of the Brout-Englert-Higgs mechanism is the only block of the standard model (SM)~\cite{SM1,SM2,Higgs1,Higgs2} whose existence has not been verified experimentally. Therefore its search is one of the most important aspects of the LHC program. The CMS design was optimized to carry out this search in
the whole allowed range and in many different production and decay channels.
The first results based on the full 2011 7 TeV center-of-mass energy data sample have been presented at the end of 2011 and submitted for publication at the beginning of 2012. We present here updated results with a more optimized analysis in the $\gamma\gamma$ channel, the addition of a few more analyses and a new combination of the results from all search channels.

The Higgs boson has been ruled out at 95\% confidence level (CL) at LEP~\cite{leplimits} with mass smaller than 114.4 GeV and at Tevatron~\cite{tevatronlimits,tevatronlimitsmoriond} with a mass in the vicinity of 160 GeV .
Indirect constraints from precision electroweak measurements favour a low mass Higgs boson~\cite{Baak:2011ze} 
above the LEP limit and give the upper limit 
%%%: $\MH<169$ GeV at 95\% CL from the standard fit and 
$\MH<143$ GeV at 95\% CL, including direct searches before LHC.

The dominant production mode at LHC is the gluon-gluon fusion followed by the vector boson fusion (VBF) and associated production with a vector boson (VH), each of which contributes less than 10\% of the total production cross section. The decay branching ratios of the Higgs boson vary with its mass and are dominated by bb and $\tau\tau$ at low mass and by WW and ZZ above 135 GeV. The $\gamma\gamma$ decay channel is important in the mass range below 130 GeV and its branching ratio is of the order of $10^{-3}$.
The values of cross section and branching ratios used in the following are taken from the LHC cross section working group~\cite{Dittmaier:2012vm,Dittmaier:2011ti}.

In 2011 we had an excellent performance of both LHC and CMS and this allowed us to collect approximately 5 \fbinv of data that are good for all analyses.
The CMS detector is a multipurpose detector and is extensively described in~\cite{CMSdetector}.
The average pileup was about 10 events per bunch crossing, and special care was taken to mitigate its effects on the analysis.
One example is the usage the particle flow algorithm that combines all detector signals in an optimal way.

%%%%%%%%%%%%%%%%%
%
% Analysis strategy
%
%%%%%%%%%%%%%%%%%

\section{Analysis strategy}

The SM scalar boson search is carried out in the mass range from 110 and 600 GeV. 
The search channels, as well as their optimization vary as function of the Higgs mass.
Table~\ref{tab:channelsnew} reports all 11 independent channels that are used for the combined results of the search, along with the corresponding integrated luminosity, the number of subchannels, the investigated mass range and the approximate Higgs boson mass resolution.
The sensitivity of the different channels varies as function of the Higgs boson mass. 
The most sensitive channel at low mass, below approximately 130 GeV, is the $\gamma\gamma$
channel. Between 130 and 200 GeV the WW channel is most sensitive, and above 200 GeV the various ZZ channels take over with ZZ$\rightarrow\ell\ell\nu\nu$ being the most sensitive in the very high mass range, above about 300 GeV.

\vspace{-0.3cm}

\begin{table}[htbp]
\begin{center}
\small
  \caption[ ] {The 11 Higgs boson search channels. The most relevant information is indicated for each of the analyses.
  }
  \label{tab:channelsnew}
\begin{tabular}{ l c c c c c}
\hline %-----------------------------------------------------------------------------------------------------
\hline %-----------------------------------------------------------------------------------------------------
\multirow{2}{*} {Channel}                            & $m_H$ range      & Luminosity     & Sub-      & $\mH$        &   \multirow{2}{*} {Comment}   \\
                                                     & (\GeV)           & (fb$^{-1}$)    & channels  & resolution           \\
\hline %-----------------------------------------------------------------------------------------------------
 {$\PH \to \Pgg\Pgg$                                     } & 110--150      & 4.8         &  2        & 1--2\%    & updated       \\
    $\PH \to \Pgt\Pgt \to e\tau_{\mathrm{h}}/\mu\tau_{\mathrm{h}}/e\mu+X$  & 110--145      & 4.6         &  9        & 20\%   &           \\
 $\PH \to \Pgt\Pgt \to \mu\mu+X$                                        & 110--140      & 4.5         &  3        & 20\%     & new             \\
 $W\PH \to e\mu \tau_h / \mu\mu \tau_h + \nu$'s                         & 100--140      & 4.7         &  2        & 20\%     & new              \\
    $(\PW/\cPZ)\PH \to (\ell\nu/\ell\ell/\nu\nu)(\cPqb\cPqb)$         & 110--135      & 4.7         &  5        & 10\%        &    \\
    {$\PH \to \PW\PW^* \to 2\ell 2\nu$                      } & 110--600      & 4.6         &  5        & 20\%        &   \\
 $W\PH \to \PW(\PW\PW^*) \to 3\ell 3\nu$                                & 110--200      & 4.6         &  1        & 20\%     & new            \\
    {$\PH \to \cPZ\cPZ^{(*)} \to 4\ell$                      } & 110--600      & 4.7         &  3        & 1--2\%      &     \\
    $\PH \to \cPZ\cPZ^{(*)} \to 2\ell 2q$  & $\left\{ \begin{array} {l} \textrm{130--164} \\ \textrm{200--600} \end{array} \right. $  & 4.6  &  6     & $ \begin{array}{l}  3\% \\ 3\% \end{array}$ &
 \\
    $\PH\to\cPZ\cPZ \to 2\ell 2\tau$                                       & 190--600      & 4.7         &  8        & 10--15\%    &   \\
    {$\PH\to\cPZ\cPZ \to 2\ell 2\nu$                     } & 250--600      & 4.6         &  2        & 7\%         &    \\
\hline %-----------------------------------------------------------------------------------------------------
\hline %-----------------------------------------------------------------------------------------------------
\end{tabular}
\end{center}
\end{table}
\vspace{-0.5cm}

\section{Low mass channels}

%%%%%%%%%%%%%%%%%
%
% Gamma gamma
%
%%%%%%%%%%%%%%%%%

\subsection{$H \to \gamma\gamma$ channel}

\begin{figure}[htbp]
  \begin{center}
    \begin{tabular}{cc}
      \vspace{-0.4cm}
      \resizebox{6.9cm}{!}{\raisebox{4.00cm}{\includegraphics{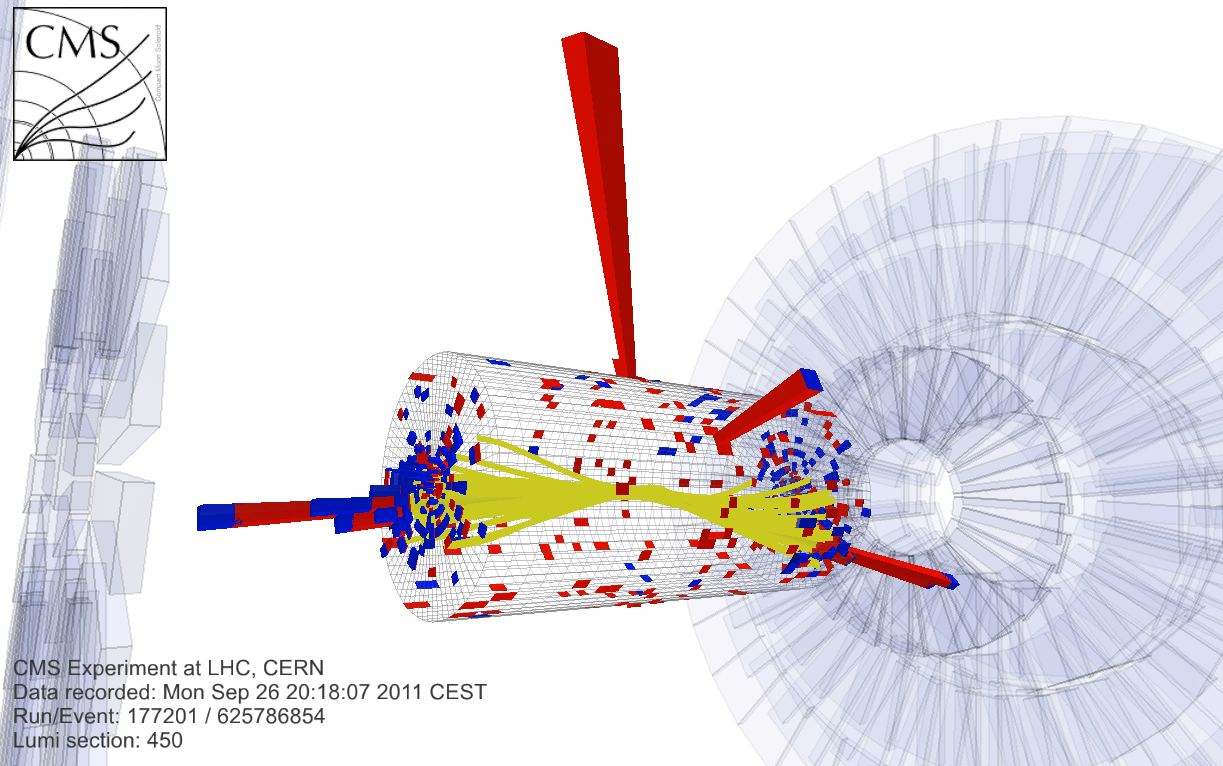}}} 
&
      \resizebox{7.5cm}{!}{\includegraphics{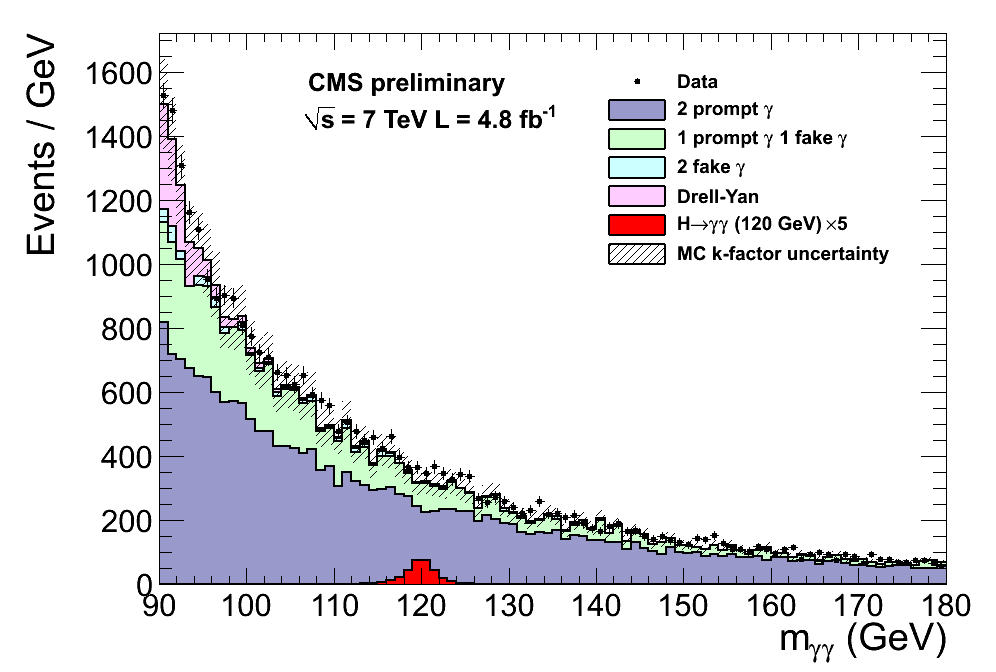}}
    \end{tabular}

    \caption{Left: VBF $\gamma\gamma$ candidate event display, Right: di-photon mass spectrum for all events passing the final selection. Data are shown together with the background MC prediction. The hatched area indicates the systematic error on the background normalization from the K-factors. The expected Higgs signal at 120 GeV is also shown superimposed and scaled by a factor 5.}
    \label{fig:gamgammassspectrum}
  \end{center}
      \vspace{-0.3cm}
\end{figure}

The Higgs boson branching ratio for the decay into two photons is approximately $2\times 10^{-3}$ between 110 and 150 GeV.
The diphoton mass resolution is very good, between 1 and 2\% and the signature in this channel is two high \et isolated photons.
In case of the VBF there are two additional high \pt jets that provide a further handle to discriminate the signal from the background.
A signal in this channel would appear like a small, narrow peak above a large and smooth background.
Figure~\ref{fig:gamgammassspectrum} shows a VBF candidate and the mass spectrum of the data and the Monte Carlo background with a superimposed Higgs signal at 120 GeV. The signal is multiplied by 5 to increase its visibility.
As can be seen from the figure, after the final selection, the background is dominated by the irreducible two photon QCD production. However there is also a relevant contribution from events in which at least one of the two identified photons is a jet faking a photon.
The MC background estimation has large uncertainties, but it enters the analysis only to help the optimization process. It is not used for the derivation of the results for which only the data and the signal MC are employed.

VBF events are selected by using the same photon identification as for the inclusive analysis described later, slightly increasing the asymmetry on the photon \et cuts and finally applying additional requirements on jet variables. 
The signal to background ratio in the di-jet tag class is relatively large, and we obtain an improvement on the exclusion sensitivity of approximately 10\% in cross section.
For the remaining events, in the analysis reported in~\cite{Chatrchyan:2012tw,HIG-12-001} the sensitivity was increased by splitting the dataset into four non overlapping event classes based on the photon pseudorapidity and shower shape. In the new analysis that we present here, categories are defined in a more optimal way using a MVA based approach that results in a higher sensitivity.
Event by event mass resolution, photon Id discriminant, di-photon kinematic variables and vertex probability are combined using a boosted decision tree (BDT).
The overall sensitivity improvement of the MVA based analysis is about 20\% in exclusion cross section that corresponds to an integrated luminosity increase of more than 50\%.

Table~\ref{tab:ClassFracs} shows the number of expected signal events, the number of data events per GeV and the estimate of the resolution in all classes.

\vspace{-0.3cm}

\begin{table}[htbp]
%\color{black} 
\begin{center}
\caption{Number of selected events in different event classes,
for a SM Higgs boson signal ($m_\mathrm{H}$=120\,GeV) and for data at 120\,GeV.
%The value given for data, expressed as events/GeV, is obtained by dividing
%the number of events in a bin of $\pm$ 10\,GeV, centred at 120\,GeV, by 20\,GeV.
The mass resolution for a SM Higgs boson signal in each
event class, is also given.}
\begin{tabular}{|l|c|c|c|c|c|}
\hline
$m_\mathrm{H}$=120\,GeV &  Class 0  & Class 1  & Class 2  & Class 3  & Dijet class \\
\hline
Total signal expected events & 3.4  & 19.3  & 18.7 & 33.0  & 2.8  \\
Data (events/GeV) & 4.5  & 55.1 & 81.3 & 229.1 & 2.1  \\
\hline
Resolution FWHM/2.35 (\%)                & 0.9 & 0.9 & 1.2 & 1.7 & 1.1 \\
\hline
\end{tabular}
\label{tab:ClassFracs}
\end{center} 
\end{table}
\vspace{-0.5cm}

For the limit and significance calculation, the background is estimated by fitting to a polynomial in the full mass range ($3^{\mathrm rd}$ to $5^{\mathrm th}$ order, depending on the class).
We found that the possible bias in the background estimation is always less than 20\% of the statistical error.
As a further cross check, we carried out another MVA analysis that has a completely different approach to estimate the background, and we verified that it gives consistent results.

\begin{figure}[htbp]
  \begin{center}
    \begin{tabular}{cc}
      \vspace{-0.4cm}

      \resizebox{7.5cm}{!}{\includegraphics{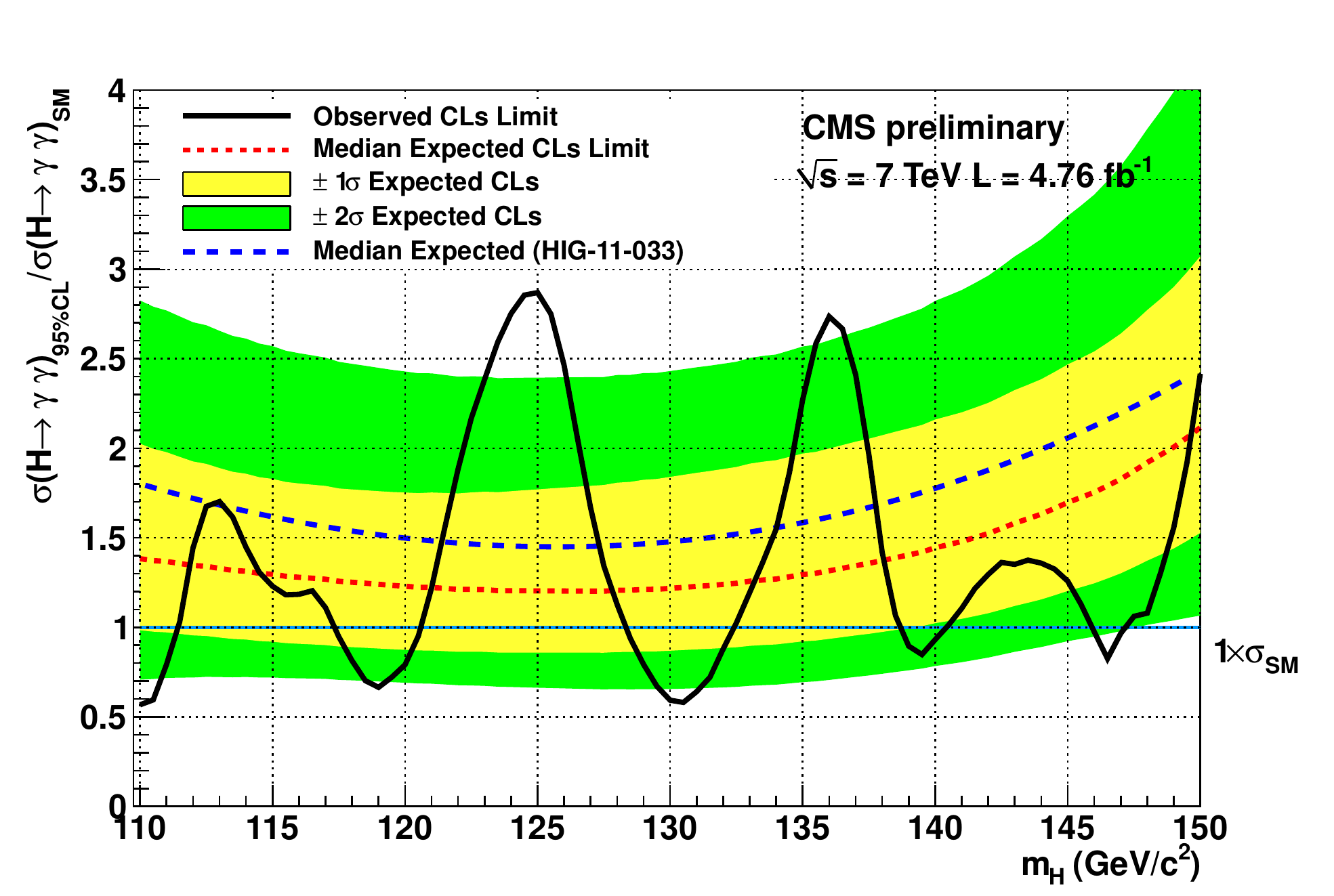}}&
      \hspace{-0.6cm} 
      \resizebox{7.5cm}{!}{\includegraphics{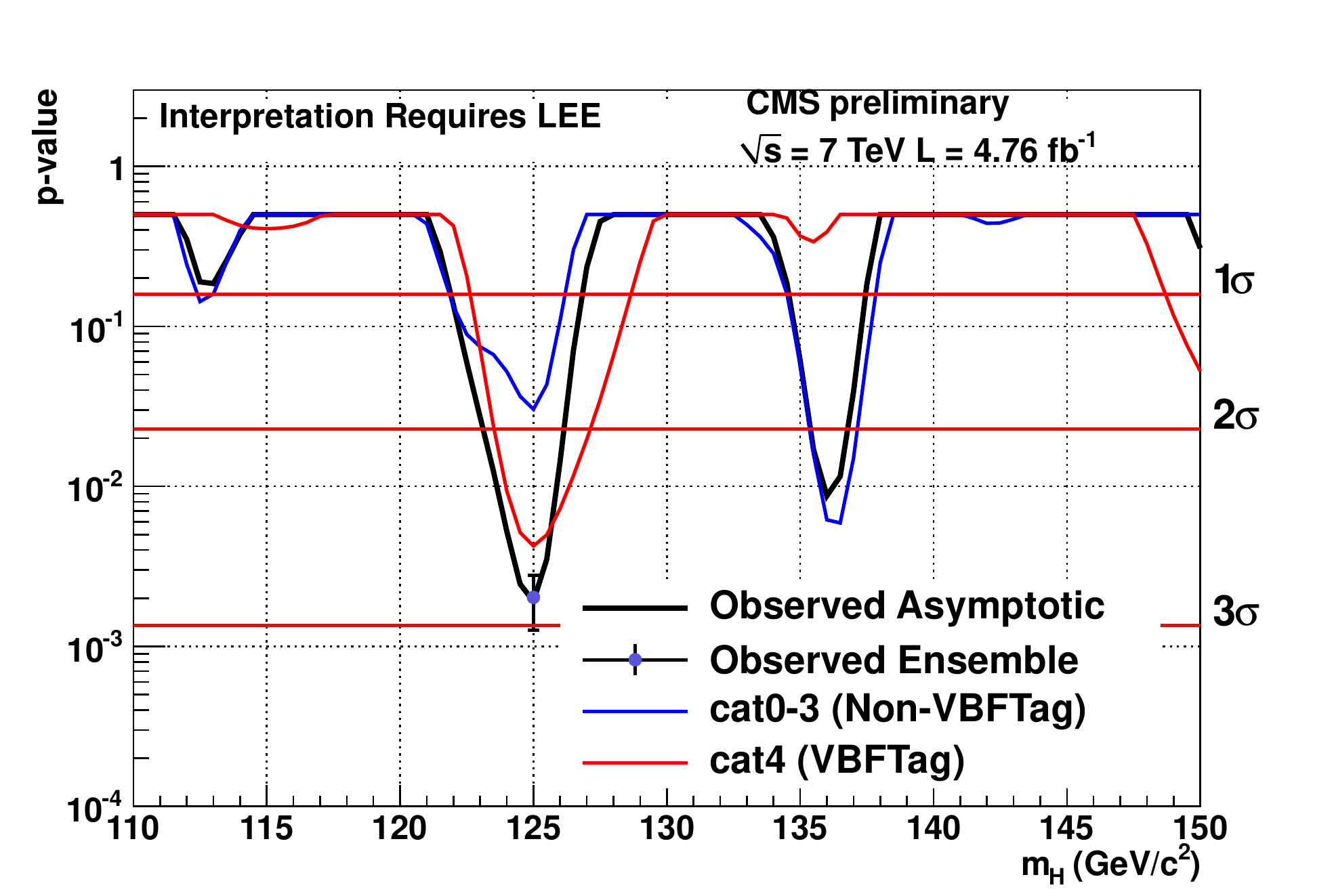}}
    \end{tabular}

    \caption{Left: 95\% exclusion on the relative signal strength to the SM in the $\gamma\gamma$ channel for the MVA based analysis. The dashed line indicates the expected limit for the cut based analysis. The yellow and green bands indicate the 1 and 2$\sigma$ expectations around the median expected result.
Right: local p-value as function of the Higgs mass. The combined p-value is shown and the VBF tag and other inclusive classes individual contributions are also shown.}
    \label{fig:combpvaluegg}
  \end{center}
      \vspace{-0.3cm}
\end{figure}

Figure~\ref{fig:combpvaluegg} shows the results in terms of 95\% CL exclusion on the cross section normalized to the SM cross section and the local p-value where the p-value is the probability that a background only fluctuation is more signal-like than the observation.
The expected 95\% CL exclusion varies between 1.2 and 2 times the SM while data exclude at 95\% CL the ranges:
110.0--111.0 GeV, 117.5--120.5 GeV, 128.5--132.0 GeV, 139.0--140.0 GeV and 146.0--147.0 GeV.
We observe the largest excess around 125 GeV with a local significance of 2.9$\sigma$. Its global significance is 1.6$\sigma$ when taking into account 
the look elsewhere effect (LEE) estimated in the full mass range 110--150 GeV.
These results are consistent with those given by the cut based analysis and the cross check MVA analysis.

%%%%%%%%%%%%%%%%%
%
% Tau tau
% b bbar
%
%%%%%%%%%%%%%%%%%

\subsection{$H \to \tau\tau$ and $H \to bb$ channels}

These two channels are the only Higgs boson decays into fermions detectable at LHC.
They are less sensitive than the $H \to \gamma\gamma$ channel, but they would be important to measure the couplings to leptons and quarks if and when the Higgs boson is discovered.
In both channels the background for the inclusive searches is huge and sensitivity is improved by requesting additional tags such as jets or charged leptons from VBF or VH production.
The $\tau\tau$ channel is also very relevant in the search for MSSM Higgs bosons.
In the case of the SM, the inclusive analysis is not very sensitive, and to improve the sensitivity we exploit 
the VBF topology as well as a boosted topology selected by requiring the presence of an additional very high \pt jet.
The mass reconstruction is not very precise due to the presence of neutrinos in the decay and the resolution is approximately 20\%.
We search in the mass range between 110 and 150 GeV~\cite{Chatrchyan:2012vp}. The expected sensitivity for exclusion is approximately 3 times the SM and we do not observe any significant excess in the data.
We have recently extended the search to cases where the both $\tau$ leptons decay into muons~\cite{HIG-12-007} and to the channel WH$\to\mathrm e\mu\tau_h, \mu\mu\tau_h$~\cite{HIG-12-006} for which we use same sign e$\mu$ and $\mu\mu$ to reduce the background from Z plus jets.
These two additional channels give another small improvement to the overall sensitivity.

In case of the Higgs decays to bb, the background from bb, produced via QCD, is much too large, so we need to
require an additional tag.
We exploit the VH associated production with W and Z decaying leptonically and we analyze separately all channels:
$e\nu$, $\mu\nu$, ee, $\mu\mu$ and $\nu\nu$~\cite{Chatrchyan:2012ww}.
We require the bb system to be boosted to improve the background rejection and the mass resolution that becomes about 10\%.
We search in the mass range between 110 and 135 GeV and the expected sensitivity for exclusion ranges from 3 to 6 times the SM.
Also in this channel we see no significant excess in the data.

The results, in terms of exclusion, are shown in Figure~\ref{fig:combcl5chan}.

\section{Channels sensitive in the full mass range}

%%%%%%%%%%%%%%%%%
%
% WW
%
%%%%%%%%%%%%%%%%%

\subsection{$H \to WW \to 2\ell 2\nu$ channel}

This is the only viable channel for the Higgs boson search around the mass region of $2\times\MW$ and the most sensitive in the mass range of approximately 125--200 GeV.
The Higgs boson mass cannot be precisely measured because of the undetected neutrinos and the resolution is of the order of 20\%.
The signature is two isolated high \pt leptons and the presence of missing transverse energy (MET). % originating from the escaping neutrinos.
The main backgrounds to this channel are WW production that is irreducible, Z plus jets, WZ, ZZ and W plus jets. The background estimation is the most important aspect of the analysis and the main backgrounds are estimated from the data.
A characteristic of the signal is that due to the fact that the Higgs boson is a scalar and to the V-A structure of the W decay, the two charged leptons tend to be aligned. This favours a small difference in azimuthal angle $\Delta\phi$ and provides some handle to discriminate the signal from the irreducible background.
 The analysis~\cite{Chatrchyan:2012ty} is performed in exclusive jet multiplicities (0, 1 and 2-jet bins) and flavour (ee, $\mu\mu$, e$\mu$) because of the different sensitivities and background contributions. For example, the irreducible WW background contributes more to the 0-jet bin, 
tt background contributes more to the 1 and 2-jet bins and 
Z plus jets and ZZ contribute more to same flavour analysis.
The 2-jet bin corresponds to the VBF analysis and again exploits the characteristics of the VBF jets such as large \pt, large $\Delta\eta$ and di-jet invariant mass.
Two types of analyses are carried out: the first is a cut-and-count for all subchannels and the second is a multivariate analysis that is applied to the 0 and 1-jet bins that are the most sensitive ones.

\begin{figure}[htbp]
      \vspace{-0.4cm}
  \begin{center}
    \begin{tabular}{cc}

      \vspace{-0.6cm}
      \resizebox{5.0cm}{!}{\includegraphics{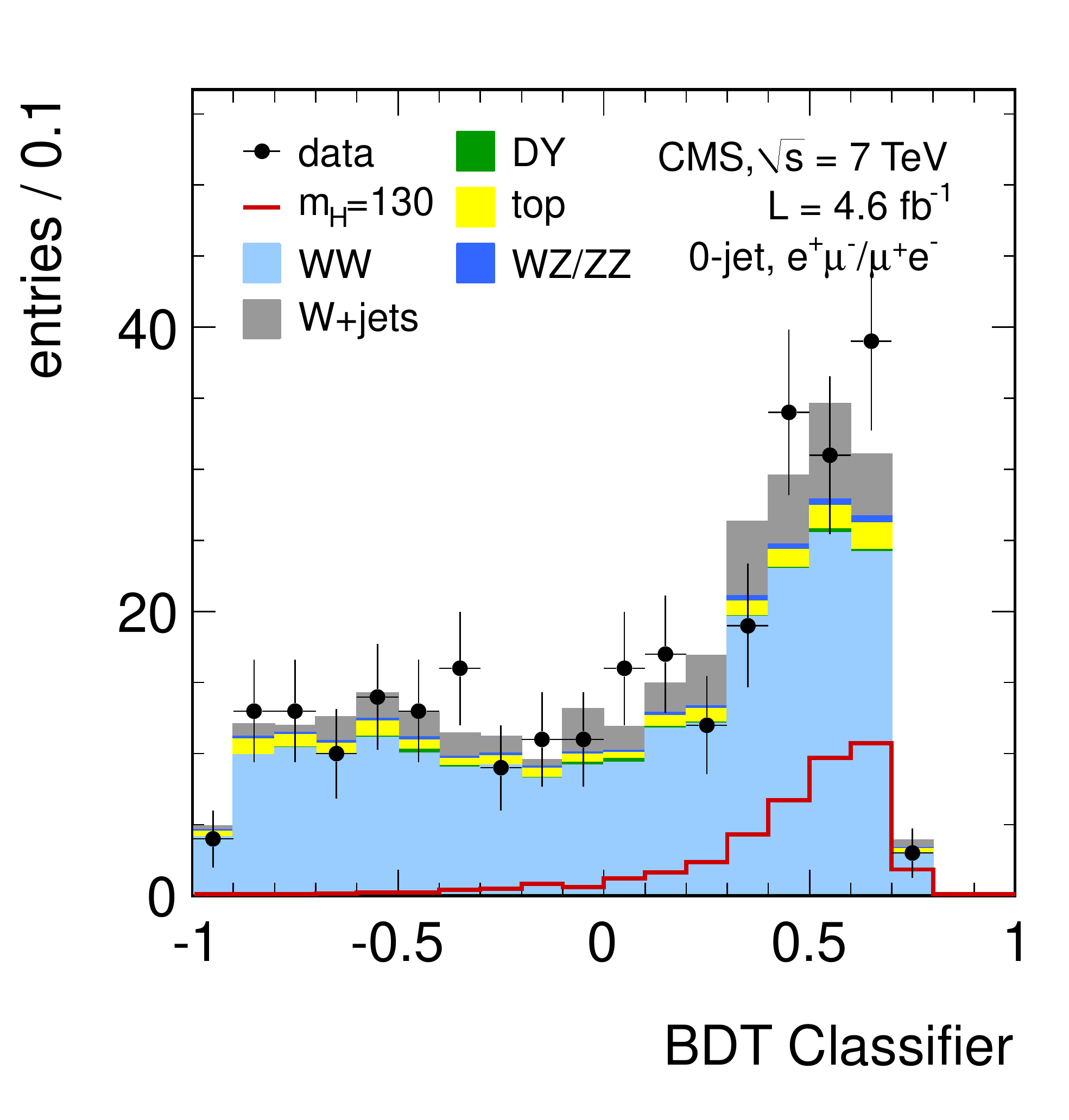}}
      &
      \resizebox{5.0cm}{!}{\includegraphics{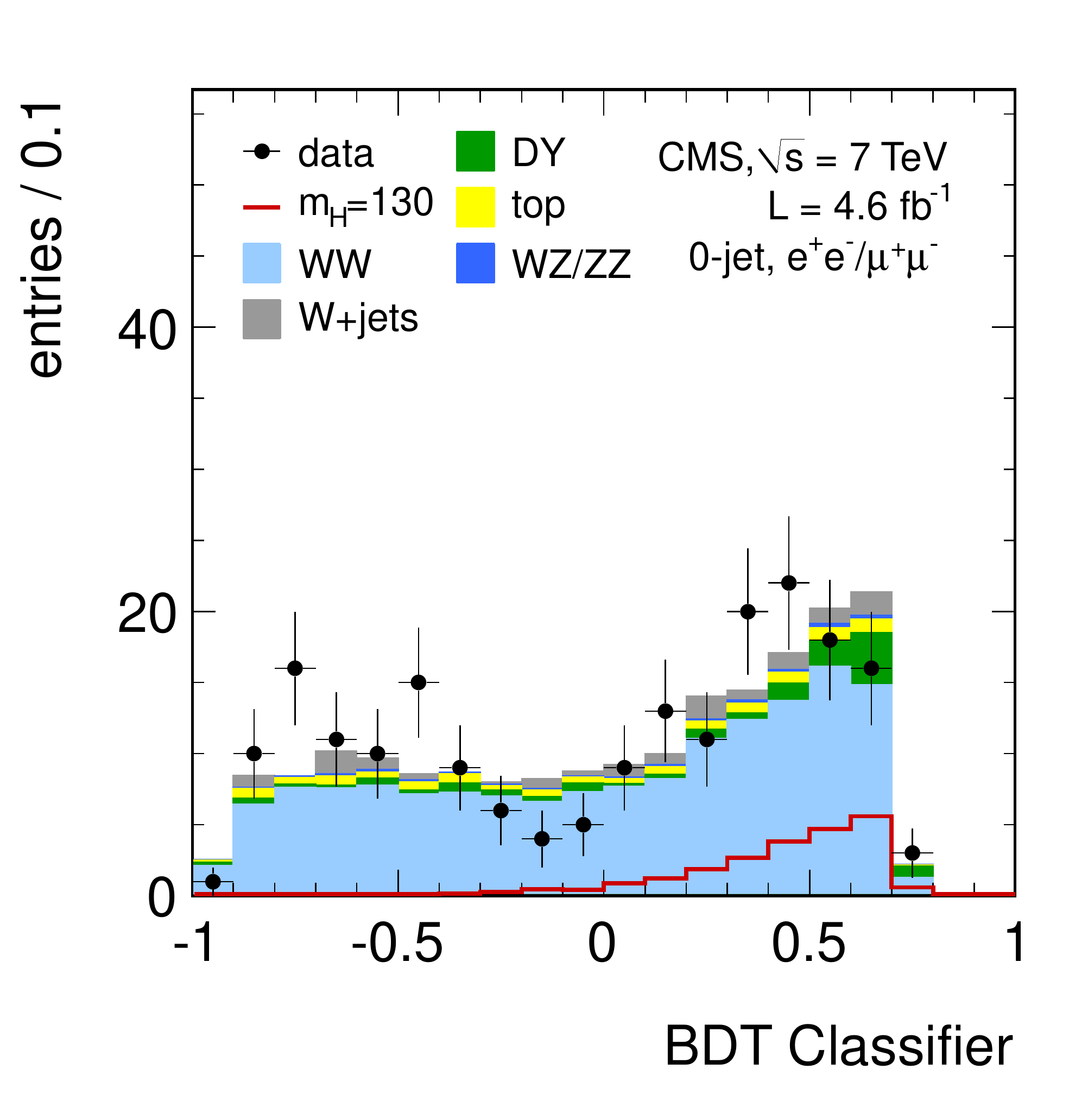}}
    \end{tabular}
    \begin{tabular}{cc}

      \vspace{-0.2cm}
\\
      \resizebox{5.0cm}{!}{\includegraphics{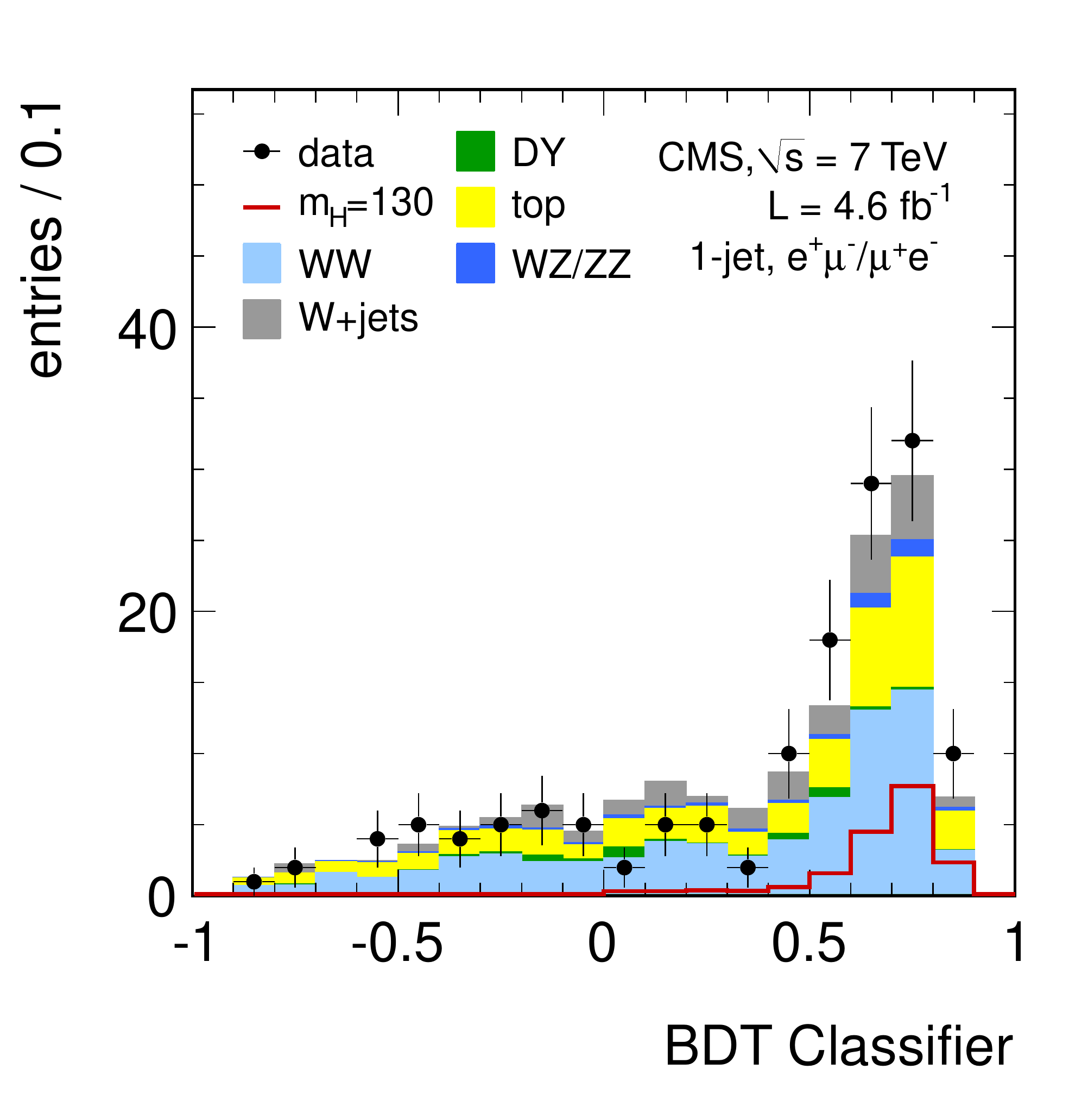}}
      &
      \resizebox{5.0cm}{!}{\includegraphics{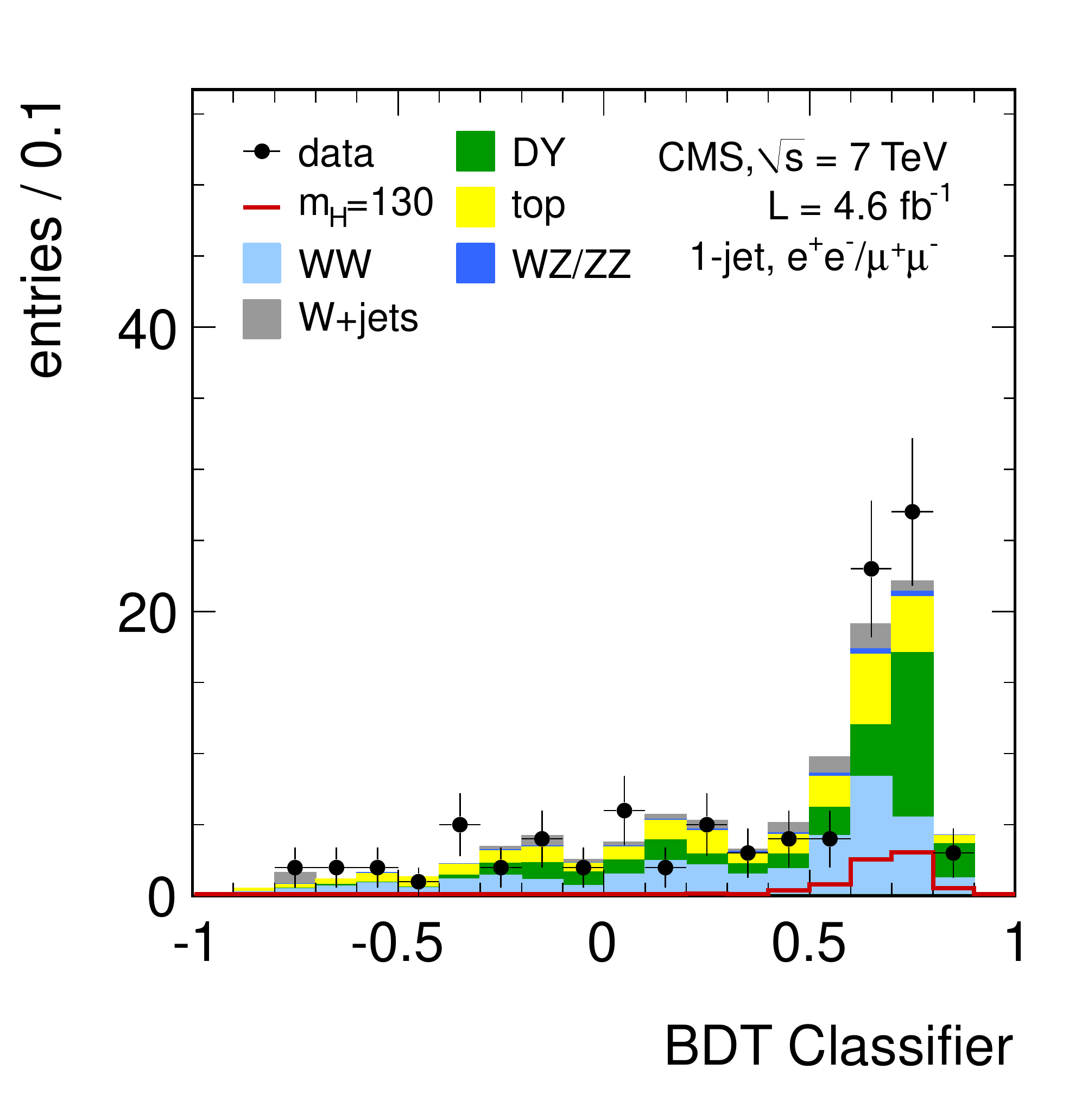}}

    \end{tabular}
      \vspace{-0.5cm}

    \caption{BDT output for the 130 GeV $H \to WW \to 2\ell 2\nu$ selection. Top left: 0-jet opposite flavour, top right: 0-jet same flavour, bottom left: 1-jet opposite flavour and bottom right: 1-jet same flavour.}
    \label{fig:wwbdtdistrib}
  \end{center}
      \vspace{-0.3cm}
\end{figure} 

The MET is a very important variable in the analysis and is affected by pileup, therefore it deserves a special treatment.
The so called projected MET is used. It corresponds to the transverse component of the MET with respect to the nearest lepton, if the angle between the MET direction and the lepton is less than $\pi/2$, and to the full MET otherwise.
Additionally, we compute two different flavours of MET: the total MET using all reconstructed particles and the charged MET using only the charged particles associated to the identified primary vertex. The minimum between the two is used for the selection. This procedure is found to reject better the background because the two definition of MET are more correlated in case of genuine MET as it is the case for the signal.
Different cuts are applied in the different flavour and same flavour channels. Cuts are tighter and a Z mass veto is applied in the same flavour channels because they are more affected by the Drell Yan background.
The cut based selection has mass dependent cuts while the MVA based analysis uses a BDT trained at different masses.
The input variables are: \pt of the leptons, ${\mathrm M}_{\ell\ell}$, $\Delta\phi_{\ell\ell}$, $\Delta\mathrm{R}_{\ell\ell}$, transverse mass of the dilepton system and of each lepton and the MET.
The overall uncertainties after the final selection are approximately 20\% for the signal efficiency and 15\% for the expected background.

Figure~\ref{fig:wwbdtdistrib} shows the final distribution of the BDT discriminant for the 0 and 1-jet bin, same flavour and opposite flavour, that is used to derive the final confidence level.
We can see that the most sensitive channel is the opposite flavour 0-jet bin where the signal is larger, the signal/background is larger and the background is dominated by the irreducible WW that has less uncertainties than the Z plus jets or tt contributions.

\begin{figure}[htbp]
  \begin{center}
    \begin{tabular}{cc}

      \vspace{-0.4cm}
      \resizebox{7.5cm}{!}{\includegraphics{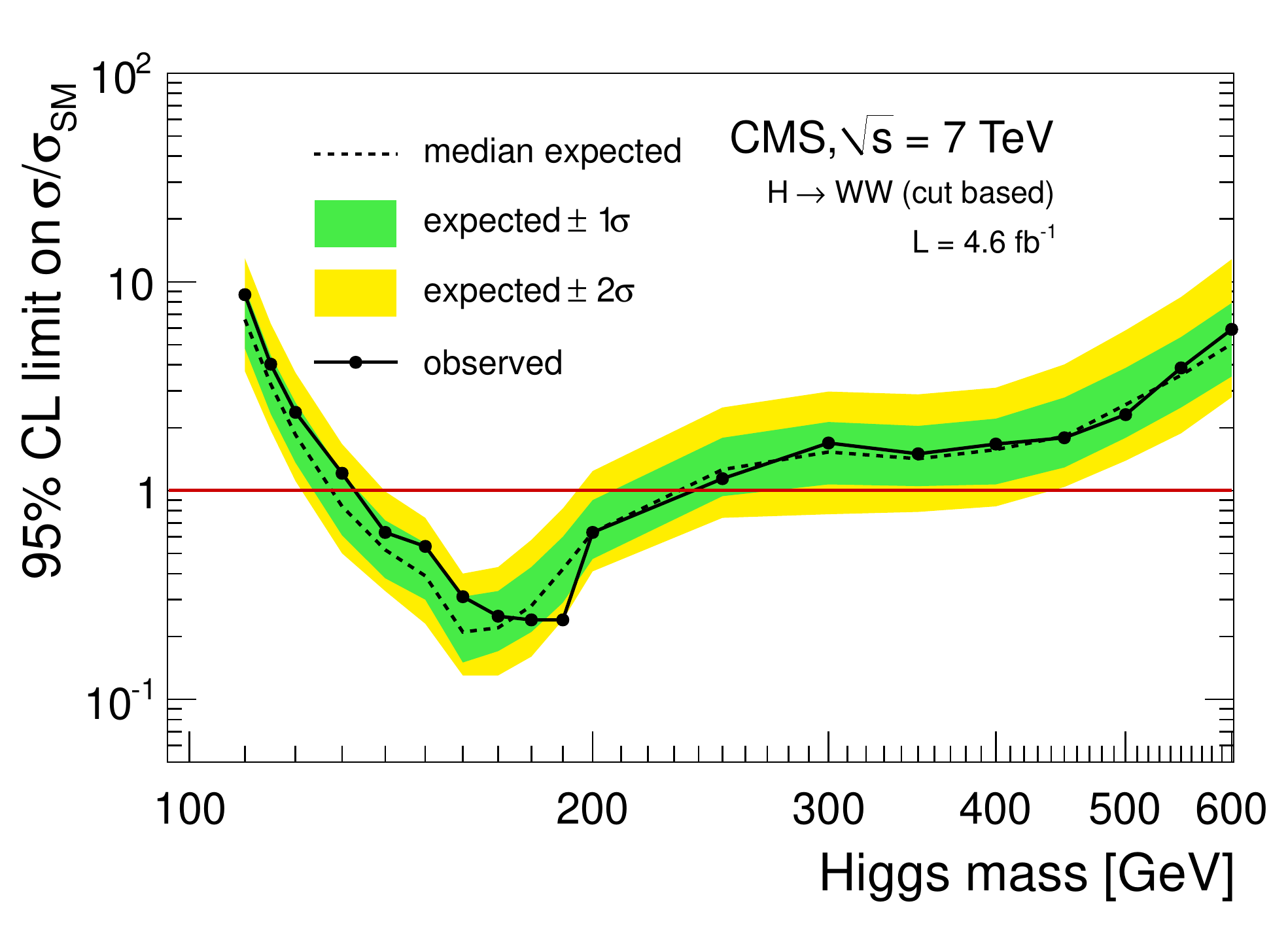}}
      &
      \resizebox{7.5cm}{!}{\includegraphics{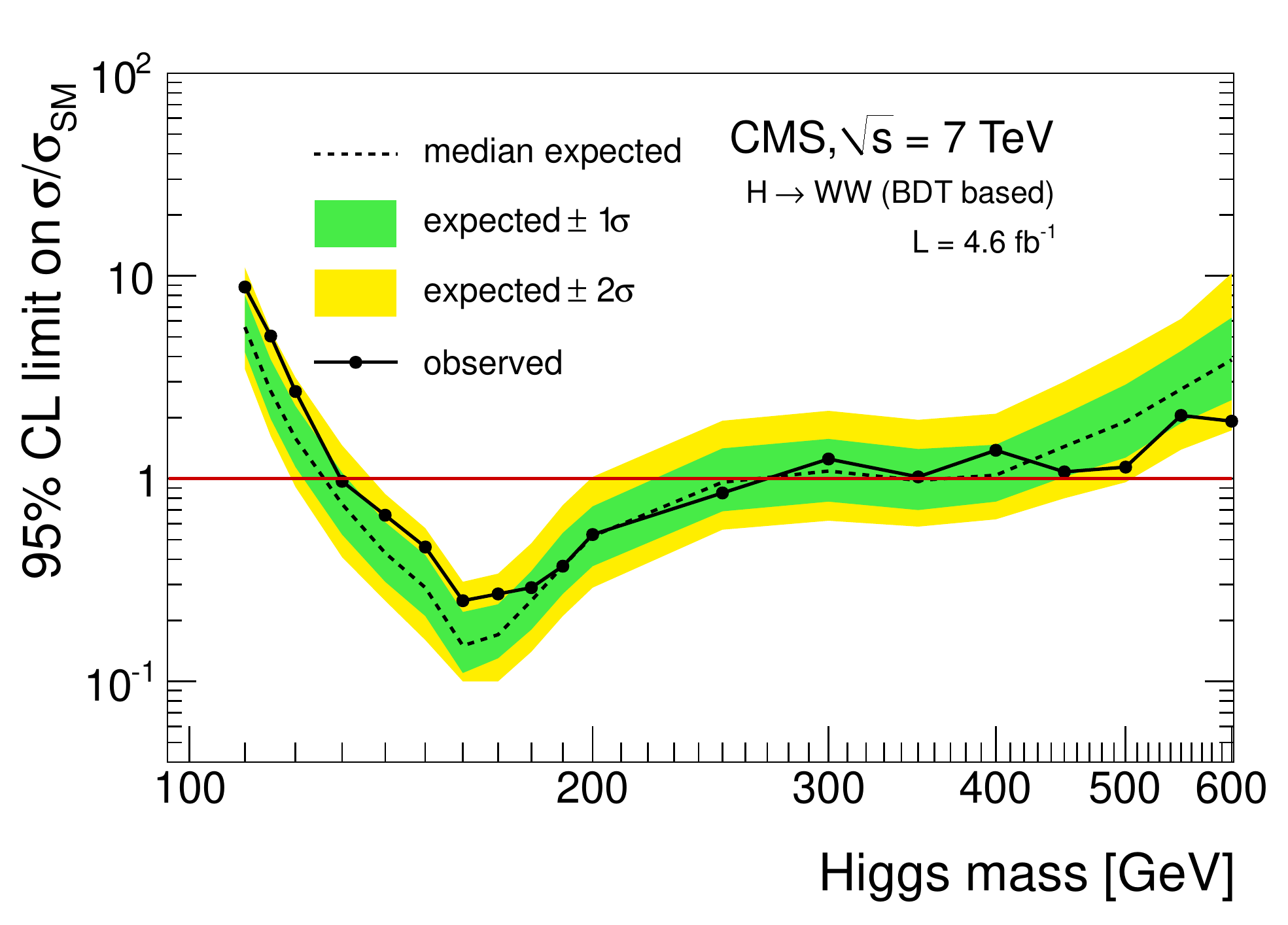}}
    \end{tabular}

    \caption{95\% exclusion limit on the relative signal strength to the SM for the cut based analysis(left) and for the MVA analysis (right) in the $H \to WW \to 2\ell 2\nu$ channel.}
    \label{fig:WWlimit}
  \end{center}
      \vspace{-0.3cm}
\end{figure}

Figure~\ref{fig:WWlimit} shows the 95\% exclusion confidence level for the cut based and the MVA shape analysis.
We observe no significant excess in the full mass range though a small excess is observed at low mass. Therefore the observed limits are similar to the expected ones.
For the MVA shape analysis the 95\% C.L. expected exclusion is for \MH\ between 127 and 270 GeV and the range 129--270 GeV is excluded at 95\% CL.

\vspace{-0.3cm}

We recently added the WH$\to$WWW$\to 3\ell 3\nu$ channel~\cite{HIG-11-034}. This analysis is very similar to the WW channel with the main backgrounds estimated from data. It is a mass independent cut-and-count analysis and it is sensitive to about 4 times the SM in the most sensitive region around $2\times\MW$.

\vspace{-0.3cm}

\begin{table}[htbp]
%\color{black} 
\begin{center}
\caption{Expected background and observed data in the full mass range and in the mass range 100--160 GeV in the $H \to ZZ \to 4\ell$ channel.}
\begin{tabular}{|l|c|c|}
\hline
Mass range &  Expected background  & Observed data\\
\hline
Full mass range & $67\pm 6$ & 72\\
$M_{4\ell}$ in 100--160 GeV & $9.7\pm 1.3$ & 13\\
\hline
\end{tabular}
\label{tab:ZZevents}
\end{center} 
\end{table}
\vspace{-0.5cm}

%%%%%%%%%%%%%%%%%
%
% ZZ
%
%%%%%%%%%%%%%%%%%

\subsection{$H \to ZZ \to 4\ell$ channel}

The $H \to ZZ \to 4\ell$ channel is the cleanest channel and it is often referred as the ``golden channel''.
The signal consists of four isolated leptons. For high mass both pairs of opposite charge and same flavour leptons are consistent with Z decays while for lower masses at least one of the pairs has lower mass.
The Higgs branching ratio for this channel is rather small, approximately one per mille at high mass and lower for masses below $2\times\MW$ but the background is very small, consisting mainly of irreducible continuum ZZ production and, to a lesser extent, Z plus jets and especially Zbb.
The mass resolution is very good and ranges between 1 and 2\%.
The \pt of the lower \pt leptons is rather small and one of the most important features of the analysis is the achievement of a very high lepton efficiency down to very low \pt. 
The analysis is carried out in the full mass range, from 110 to 600 GeV~\cite{Chatrchyan:2012dg} and the expected background and the data passing the selection are shown in Table~\ref {tab:ZZevents}.

\begin{figure}[htbp]
  \begin{center}
    \begin{tabular}{cc}

      \vspace{-0.4cm}
      %%%\vspace{-0.15cm}
      \resizebox{6.5cm}{!}{\includegraphics{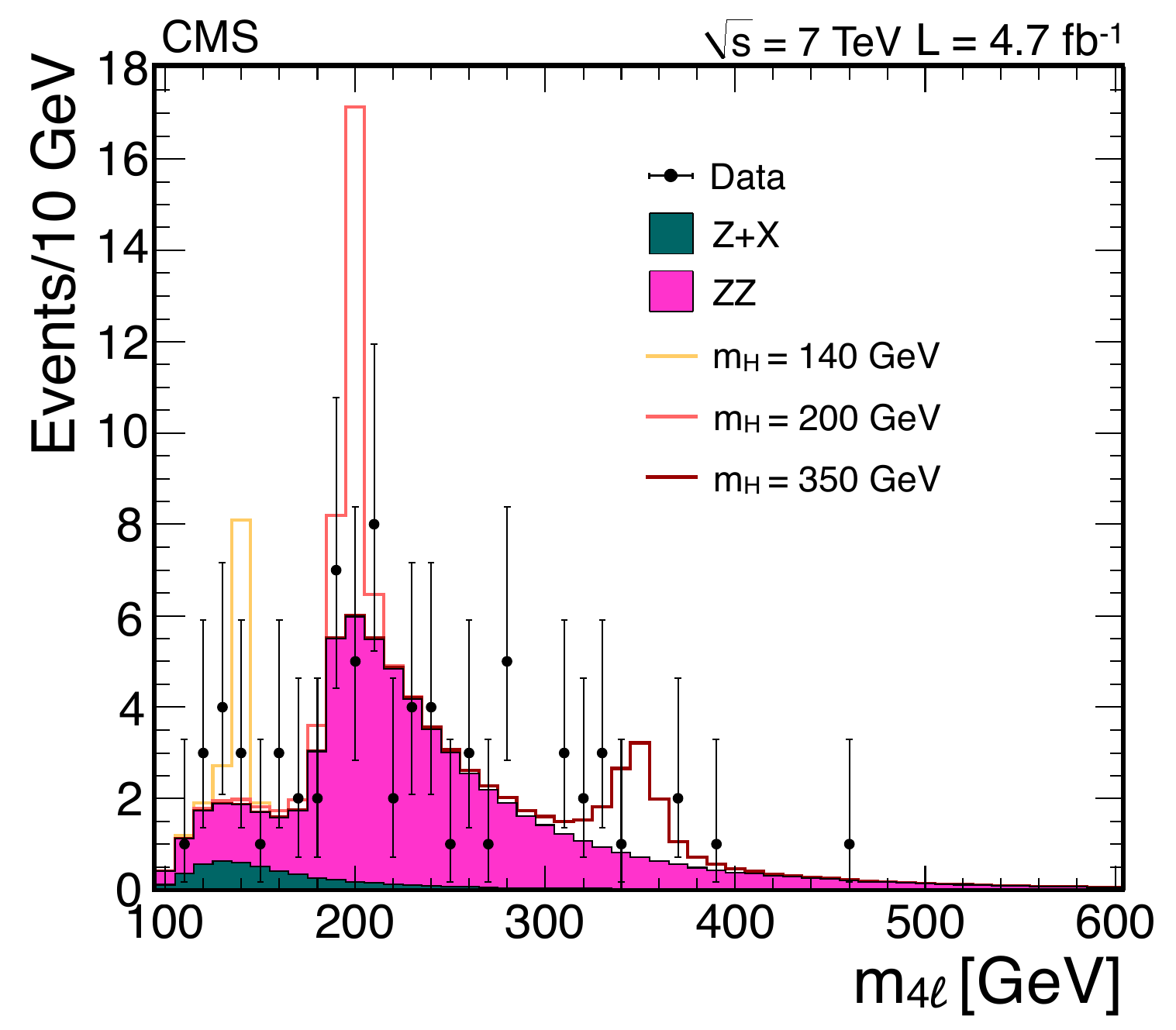}}
      &
      \resizebox{6.33cm}{!}{\raisebox{-0.33cm}{\includegraphics{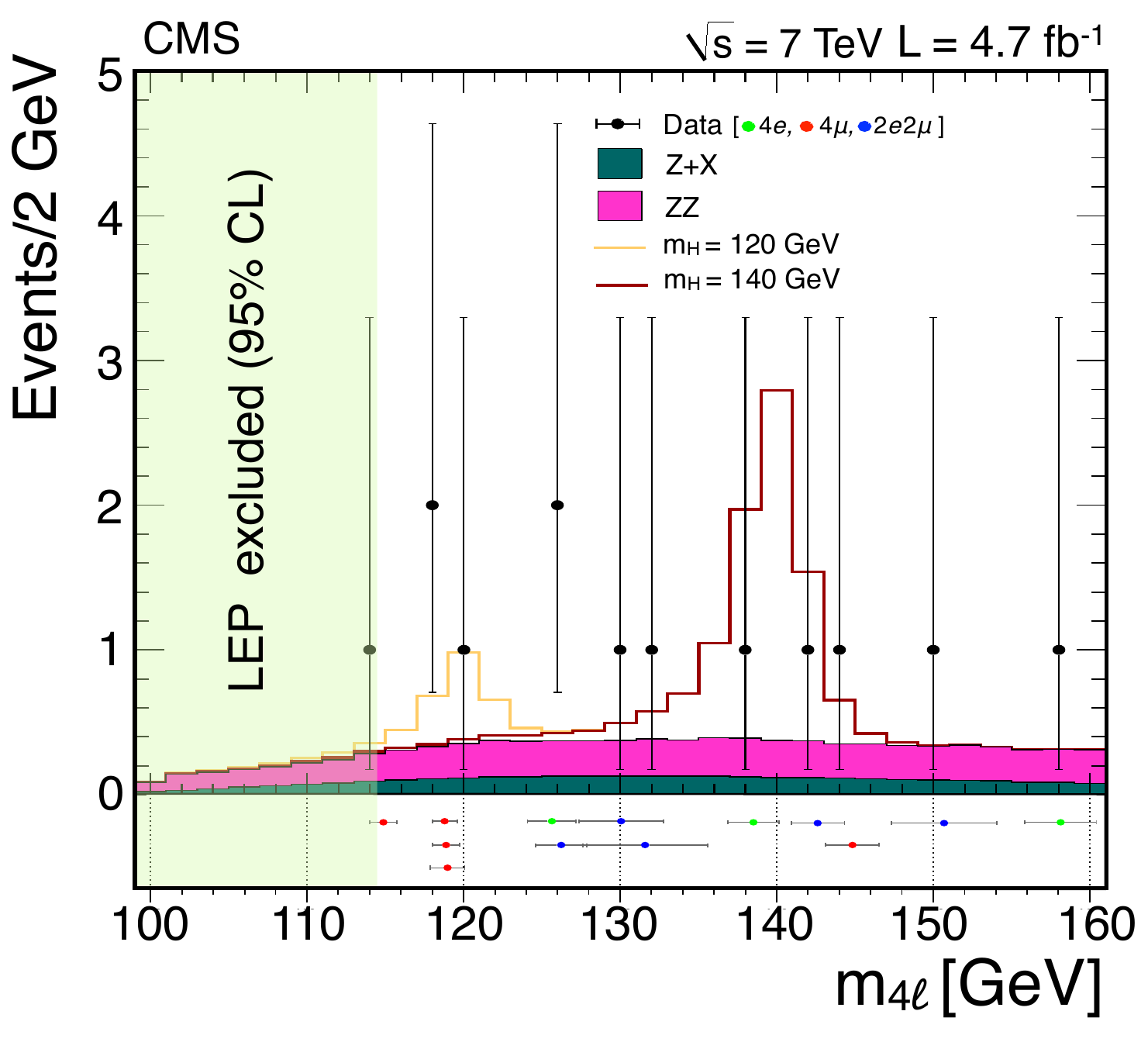}}}
    \end{tabular}

    \caption{Mass spectrum of the ZZ$\to 4\ell$ candidates in the full mass range (left) and in the low mass range (right).}
    \label{fig:ZZmass}
  \end{center}
      \vspace{-0.3cm}
\end{figure}

\begin{figure}[htbp]
  \begin{center}
    \begin{tabular}{cc}

      \vspace{-0.4cm}
      \resizebox{6.5cm}{!}{\includegraphics{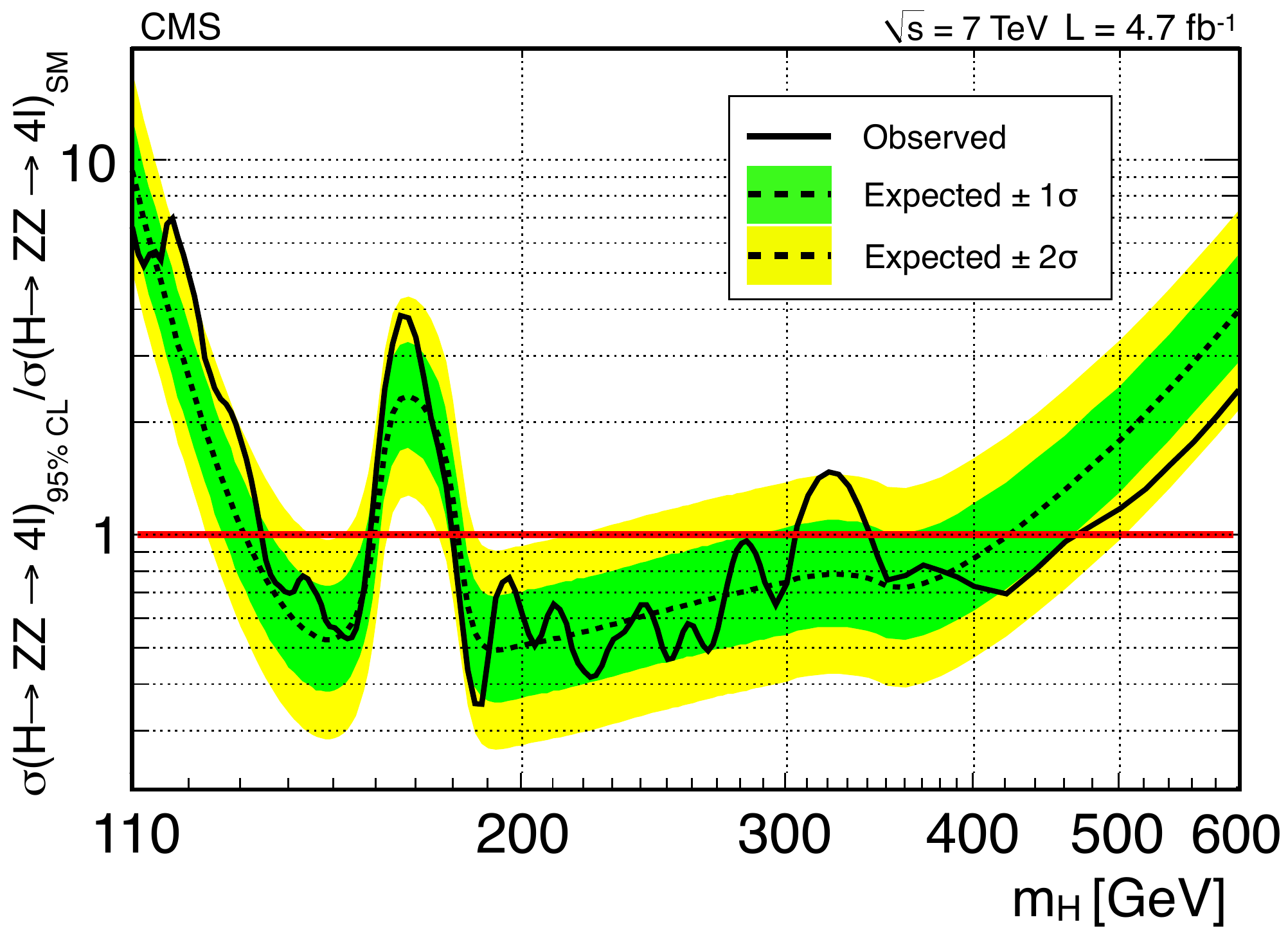}}
      &
      \resizebox{6.5cm}{!}{\includegraphics{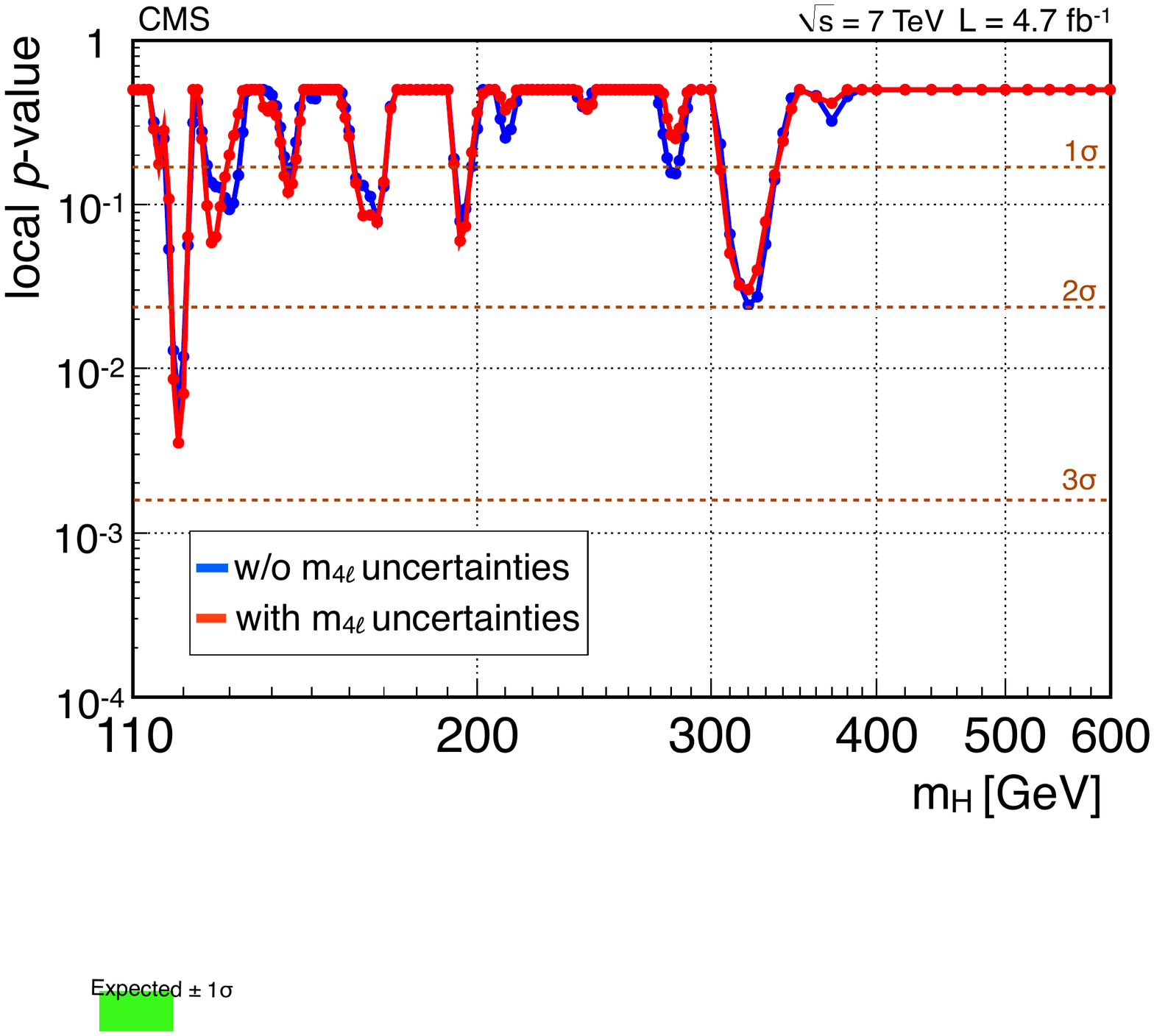}}
    \end{tabular}

    \caption{95\% exclusion limit on the relative signal strength to the SM (left) and local p-value computed with and without the individual candidate errors on the reconstructed mass in the $H \to ZZ \to 4\ell$ channel.}
    \label{fig:ZZlimpvalue}
  \end{center}
      \vspace{-0.3cm}
\end{figure} 

Figure~\ref{fig:ZZmass} shows the invariant mass spectrum of the selected data compared to the background expectations.
We do not observe any significant excess of the data and we exclude at 95\% CL
the SM Higgs boson with \MH\ in 134--158, 180--305 and 340--465 GeV.
The most significant excess is given by an accumulation of 3 events at a mass of approximately 119.5 GeV. 
It has a local significance of 2.5$\sigma$ and
a global significance of 1.0$\sigma$ in the full mass range and 1.6$\sigma$ in the mass range 100--160 GeV.

%%%%%%%%%%%%%%%%%%
%                %
%   HIGH MASS    %
%                %
%%%%%%%%%%%%%%%%%%

\section{High mass channels}

A SM Higgs boson above a mass of approximately 200 GeV almost exclusively decays into WW and ZZ and above 
about 300 GeV the Higgs boson width starts to be larger than the resolution in ZZ channels.
Beyond the previously described channels $H \to WW \to 2\ell 2\nu$ and $H \to ZZ \to 4\ell$ channels, we searched in the channels where one Z decays into $\nu$, quark and $\tau$ pairs.

\subsection{$H \to ZZ \to \ell\ell\nu\nu$ channel}

The $H \to ZZ \to \ell\ell\nu\nu$ channel has a branching ratio 6 times larger that ZZ to $4\ell$ and is the most sensitive at very high mass.
It is only accessible for high mass ($\MH>250$ GeV) because the two Z bosons need to be boosted to give rise to MET by means of the Z invisible decay.
Missing neutrinos worsen the mass resolution that is about 7\% in this channel.
The main backgrounds in this channel are again the irreducible ZZ process, Z plus jets, tt and WZ.
The background estimation is very important and the main backgrounds are estimated from data control samples. 
The background from Z plus jets is estimated using $\gamma$ plus jet events that are used to model the MET distribution.
The non-resonant background normalization is estimated from e$\mu$ events.
Again in this channel two independent analyses are carried out~\cite{Chatrchyan:2012ft}: a cut based analysis and a mass shape analysis that is more sensitive. The latter uses as 
final discriminant variable the transverse mass defined as:
$$\MT^{2} = \left(\sqrt{{\PT(\ell\ell)}^{2} + {M(\ell\ell)}^{2}} + \sqrt{{\MET}^{2} +
{M(\ell\ell)}^{2}}\right)^{2} - (\vec{p}_{T}(\ell\ell) + \vec{E}_T^\text{miss})^{2}.$$
$\PT(\ell\ell)$ and $M(\ell\ell)$ are the transverse momentum and the invariant mass of the dilepton system respectively.
We did not observe any excess in the data and the observed exclusion from this channel alone is similar to the one expected in presence of background only.
The expected  95\% CL exclusion using this channel alone is \MH\ in 290--480 GeV and the observed is \MH\ in 270--440 GeV.

\subsection{$H \to ZZ \to \ell\ell$qq and $H \to ZZ \to \ell\ell\tau\tau$ channels}

The $H \to ZZ \to \ell\ell$qq channel~\cite{Chatrchyan:2012sn} is used both for the high mass, where its sensitivity is similar but a little lower than the other ZZ channels,  and for lower masses where it only gives a small contribution to the sensitivity.
The $H \to ZZ \to \ell\ell\tau\tau$ channel~\cite{Chatrchyan:2012hr} has a lower sensitivity of about 4 times the SM.
The overall sensitivity and observed limit for all ZZ channels combined is shown in Figure~\ref{fig:combcl5chan}.

\section{Combination of all channels}
\label{sec:Results}

All searched channels are combined to obtain the final exclusion and discovery confidence levels.
The combination is carried out using the so-called CLs method described in~\cite{LHClimits}.
The combination of the published results 
is reported in 
reference\cite{Chatrchyan:2012tx}.
Here we present the combination that includes the new preliminary results presented at this conference~\cite{HIG-12-008}.
SM cross sections and branching ratios are assumed for the combination 
with their theoretical uncertainties~\cite{Dittmaier:2012vm,Dittmaier:2011ti}. An overall signal strength multiplier $\mu=\sigma/\sigma_{\mathrm{SM}}$ is introduced and limits on its value are derived.

\begin{figure}[htbp]
  \begin{center}
    \begin{tabular}{cc}
      \vspace{-0.4cm}
      \resizebox{5.0cm}{!}{\includegraphics{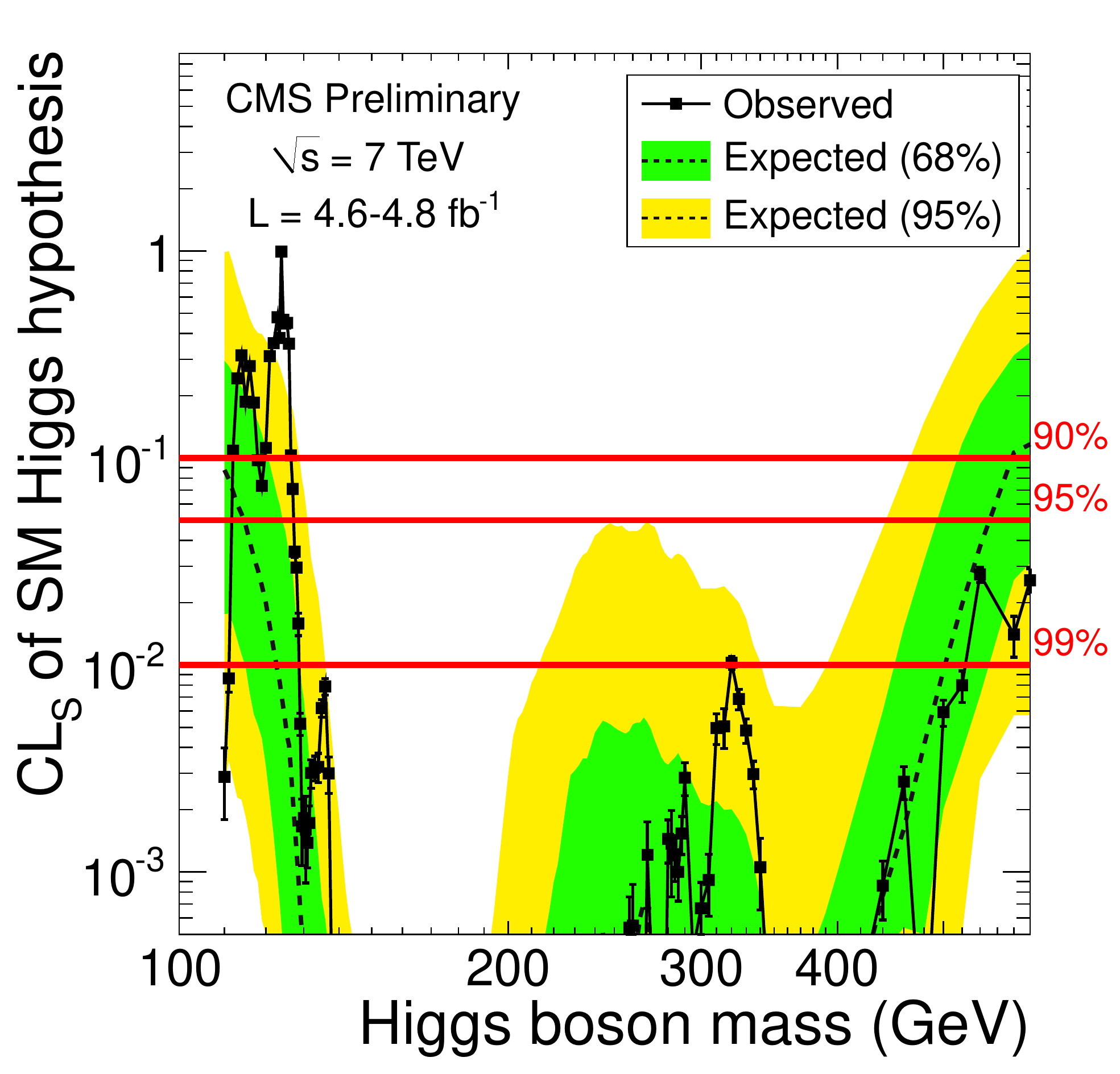}}
      \resizebox{5.0cm}{!}{\includegraphics{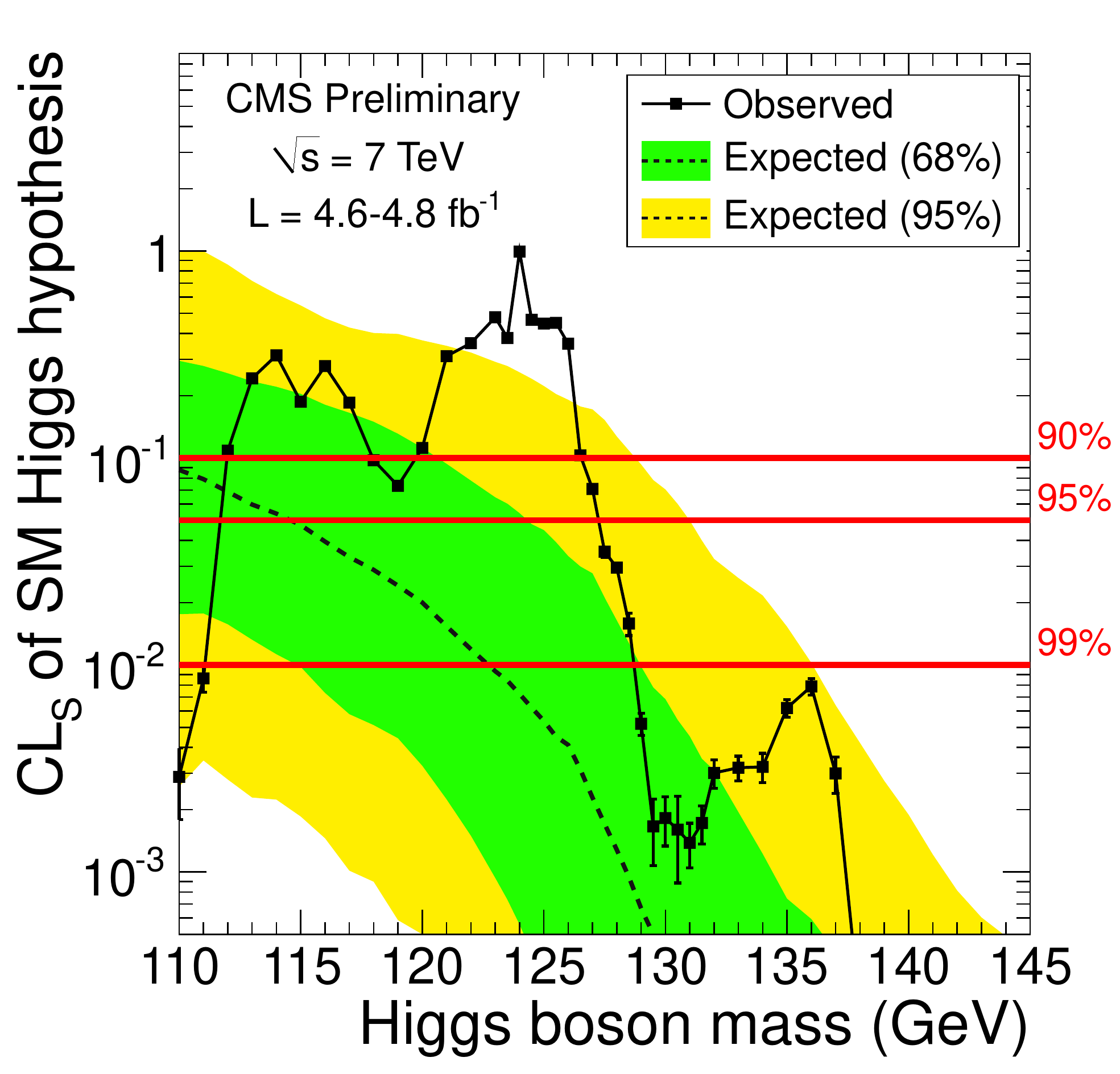}}
    \end{tabular}
    \caption{Exclusion confidence level for the combined SM Higgs search in the full mass range 110--600 GeV (left) and low mass zoom (right).
The solid line indicates the observed confidence level and the dashed line the expected one.}
    \label{fig:combcl}
  \end{center}
      \vspace{-0.3cm}

% additional
      \vspace{-0.3cm}

\end{figure} 

\begin{figure}[htbp]
  \begin{center}
    \begin{tabular}{cc}
      \vspace{-0.4cm}

      \resizebox{5.0cm}{!}{\includegraphics{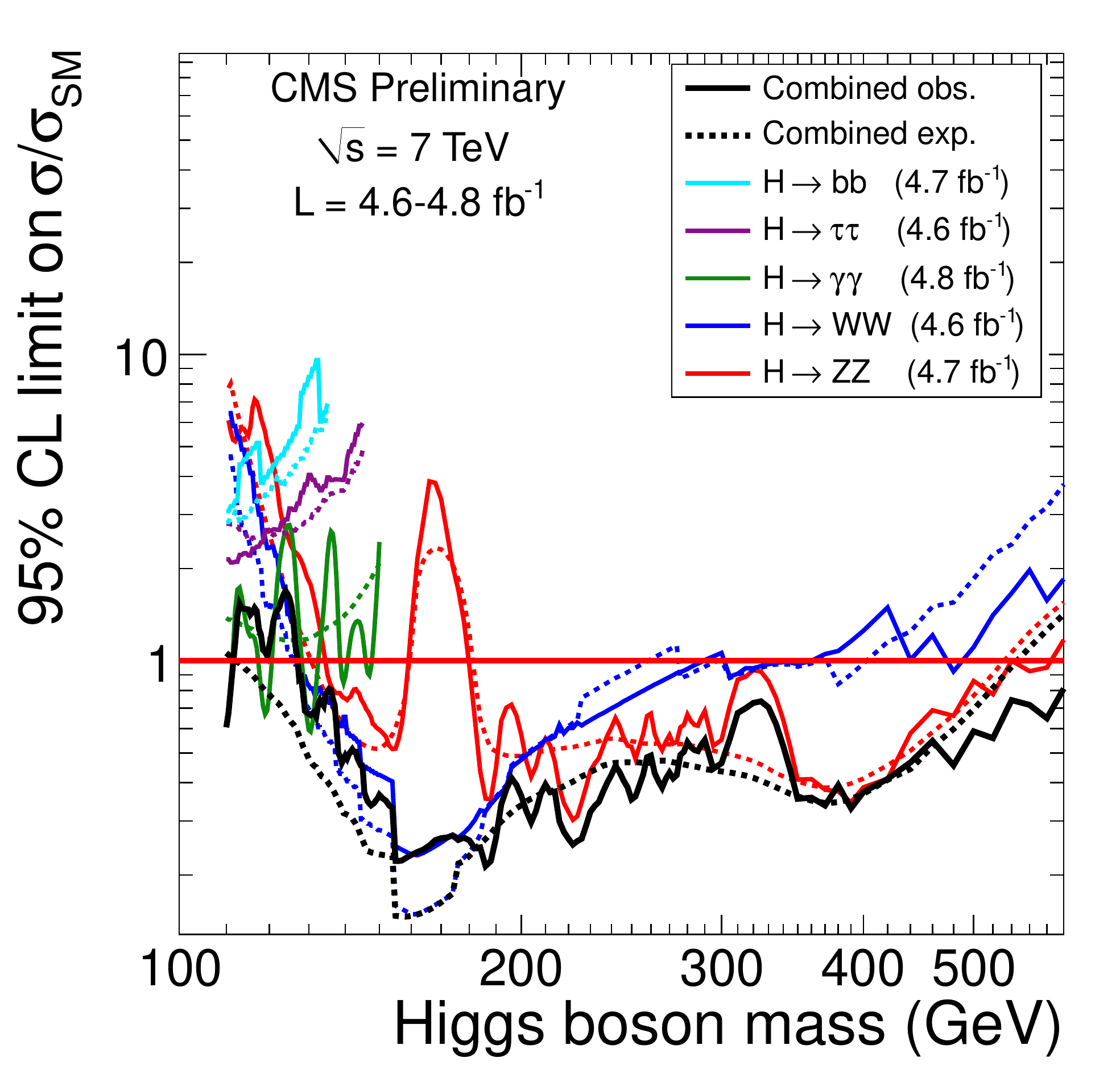}}
      \resizebox{5.0cm}{!}{\includegraphics{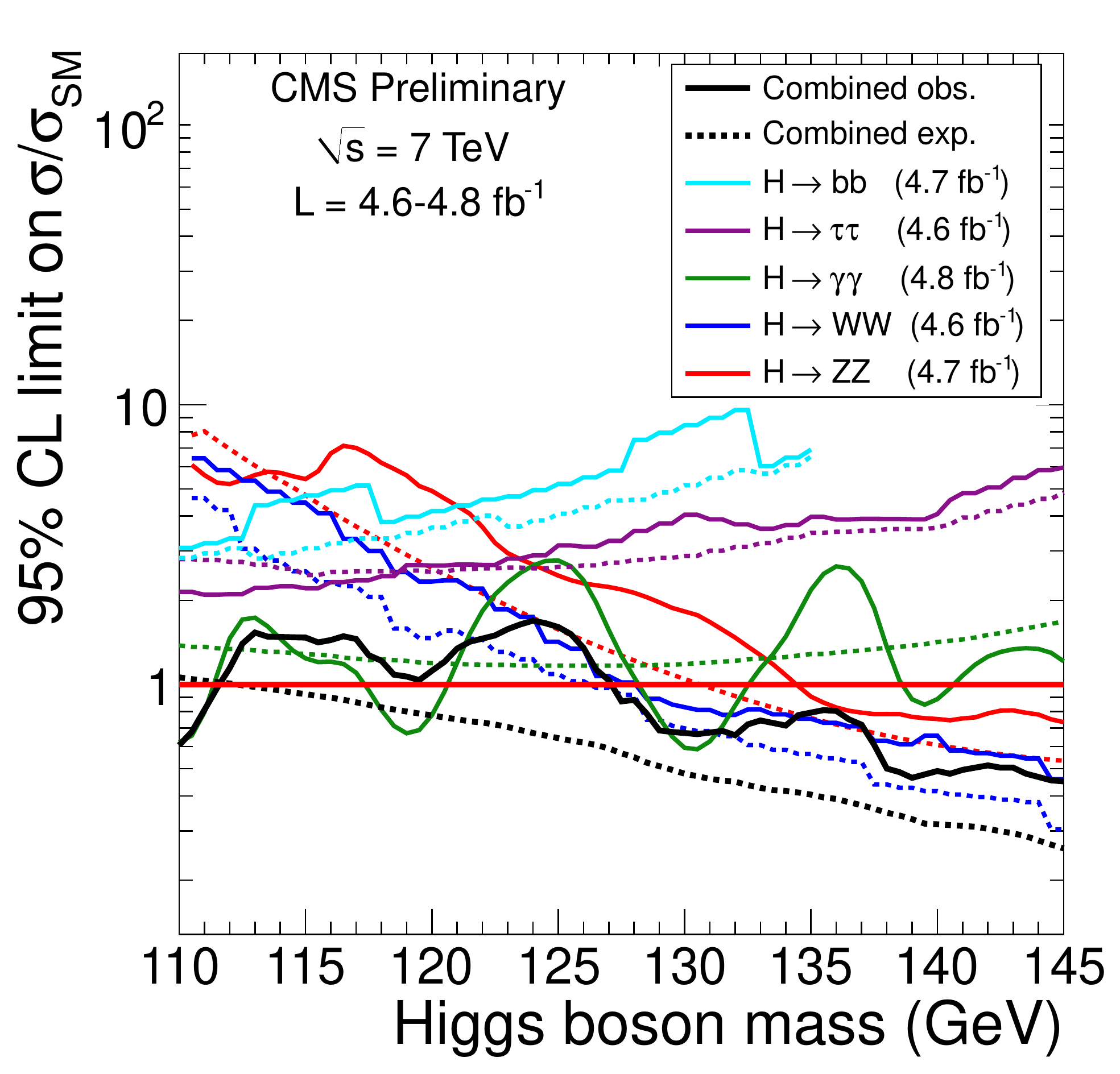}}
    \end{tabular}

    \caption{95\% exclusion confidence level on the signal strength multiplier for the SM Higgs search in the 5 Higgs decay channels.
The solid lines indicate the observed exclusion and the dashed lines the expected.}
    \label{fig:combcl5chan}
  \end{center}
      \vspace{-0.3cm}
\end{figure}

Figure~\ref{fig:combcl} shows the SM exclusion confidence level as function of the Higgs boson mass. 
The SM Higgs boson is excluded by our search at 95\% confidence level in the range 127.5--600 GeV and at 99\% confidence level
in the range 129--525 GeV.
The expected 95\% exclusion is 114.5--543 GeV and gets extremely close (100 MeV away) to the LEP lower limit.
The observed CMS upper limit on the Higgs boson mass is higher than expected in case of no signal because of the excess that is observed in the data in the region between 115 and 128 GeV.
Figure~\ref{fig:combcl5chan} shows the 95\% exclusion limit on the signal strength multiplier $\mu$ in the different Higgs decay channels.

\begin{figure}[htbp]
      \vspace{-0.3cm}
  \begin{center}
    \begin{tabular}{c}
      %\vspace{-0.5cm}

      \resizebox{8.5cm}{!}{\includegraphics{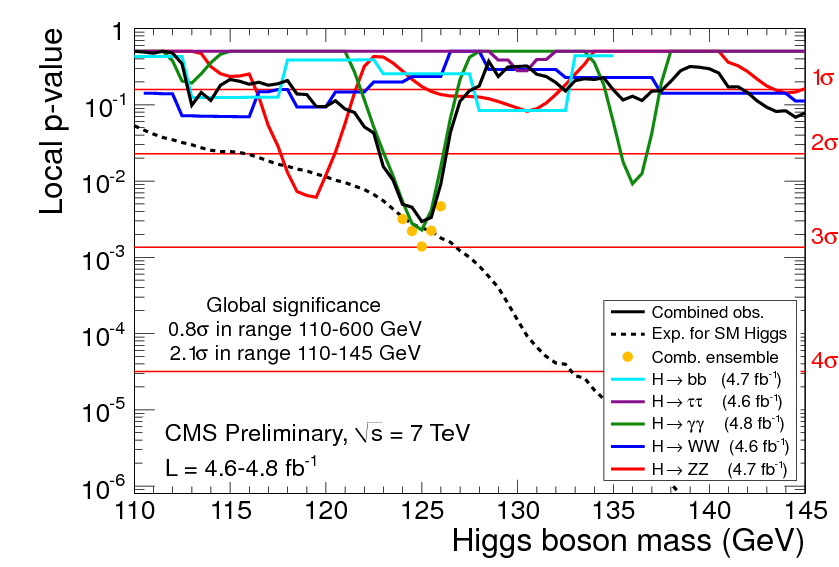}}
    \end{tabular}

    \caption{Combined local p-value for the SM Higgs search.}
    \label{fig:combpvalue}
  \end{center}
      \vspace{-0.3cm}
\end{figure}

%To quantify the significance of the observed excess we compute the p-value that indicates the probability that
%in presence of background only we obtain a more signal-like observation than the one in the data.
Figure~\ref{fig:combpvalue} shows the local p-value as function of the Higgs boson mass in the low mass region.
We can see that the minimum combined p-value is observed at a mass of 125 GeV with a local significance of $2.8\sigma$.
A similar significance is expected in presence of a 125 GeV Higgs boson signal.
If we consider the probability of observing
a local significance smaller than $2.8\sigma$ anywhere in the search range, we obtain a global significance of $0.8\sigma$ relative to the full mass range 110--600 GeV and of $2.1\sigma$ for the mass range 110--145 GeV.

\begin{figure}[htbp]
  \begin{center}
    \begin{tabular}{cc}

      \vspace{-0.6cm}
      \resizebox{6.5cm}{!}{\includegraphics{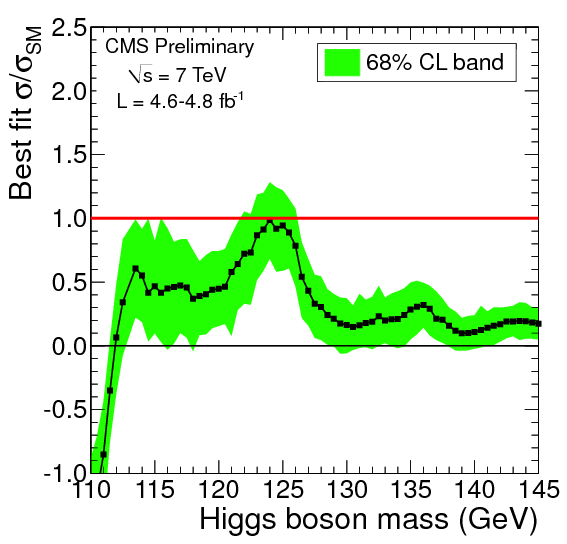}}
      &
      \resizebox{6.5cm}{!}{\includegraphics{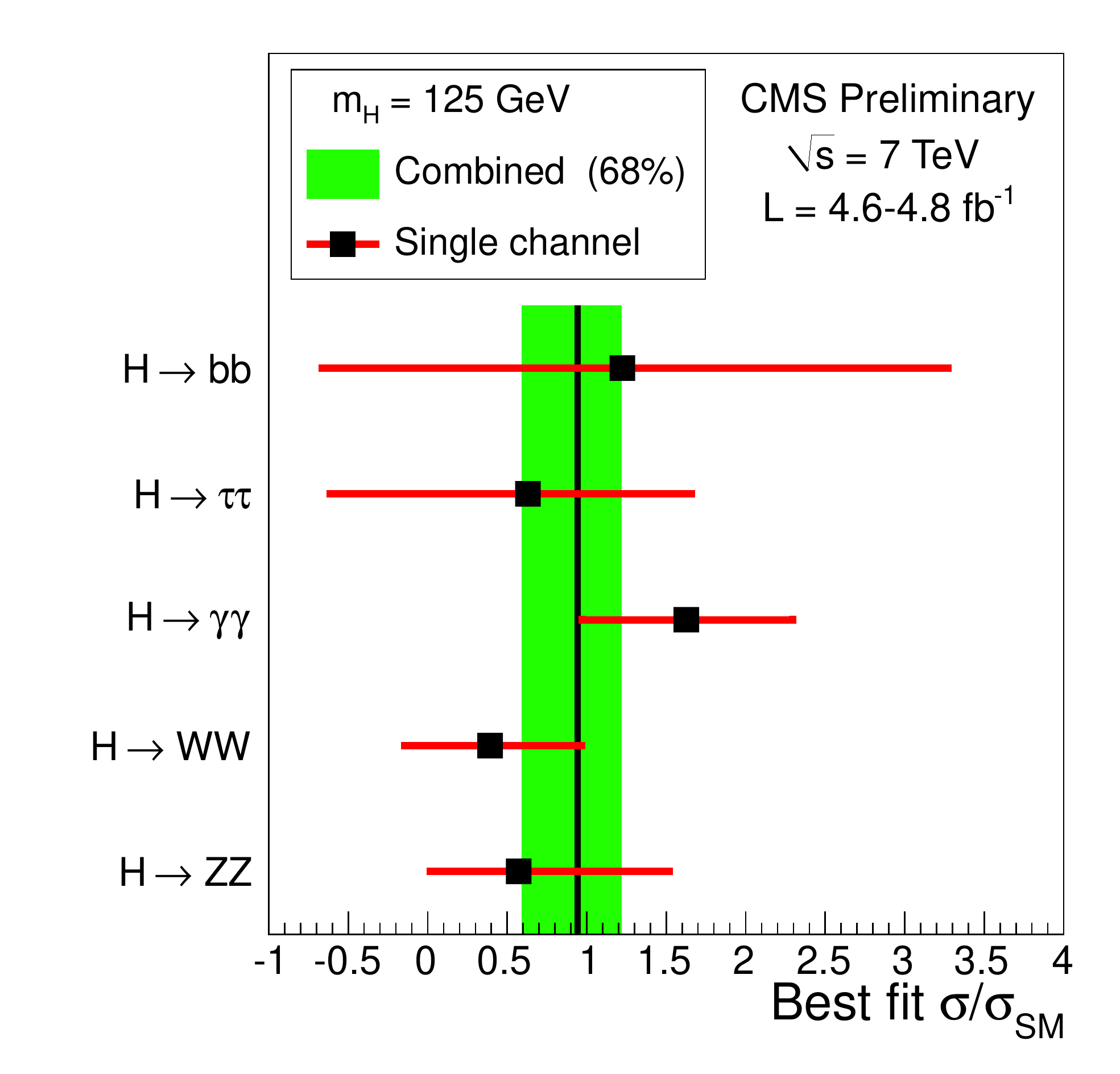}}
    \end{tabular}

    \caption{Fitted signal strength multiplier $\mu$ as function of the Higgs mass (left) and measured $\mu$ in the different channels at \MH = 125 GeV (right).}
    \label{fig:CombMu}
  \end{center}
      \vspace{-0.3cm}
\end{figure} 

Figure~\ref{fig:CombMu} shows the fitted value of the signal strength multiplier $\mu=\sigma/\sigma_{\mathrm{SM}}$ as function of the Higgs mass 
and the individual fitted values at \MH=125 GeV in the 5 sensitive channels.
The fitted $\mu$ of the excess near 125 GeV is consistent with the SM scalar boson expectation and
several channels show some excess, though most of it comes from the $H \to \gamma\gamma$ channel.
More data are needed to investigate this excess.

\section{Summary}

We searched for the SM Higgs boson in 11 independent channels using approximately 5 \fbinv of 7 TeV pp collision data collected with the CMS detector at LHC.
Combining the results of the different searches we exclude at 95\% confidence level a SM Higgs boson with mass between 127.5 and 600 GeV. 
The expected  95\% confidence level exclusion if the Higgs boson is not present is from 114.5 and 543 GeV. 
%%%  SM scalar boson if it exists, is limited at 95% CL in the range [114.4-­‐127.5] GeV
The observed exclusion is weaker than expected at low mass because of some excess that is observed below about 128 GeV. 
The most significant excess is found at 125 GeV with a local significance of $2.8 \sigma$. It has a global significance of $0.8 \sigma$ when evaluated in the full search range and of $2.1 \sigma$ when evaluated in the range 110--145 GeV. 
The excess is consistent both with background fluctuation and a SM Higgs boson with mass of about 125 GeV and more data are needed to investigate its origin.
The data that will be collected in 2012 at 8 TeV CM energy should allow us to discover or exclude the SM scalar boson.

\end{document}